\documentclass[12pt]{article}
\usepackage{epsfig}
\begin{document}
\title{Robust Hadamard gate for optical and ion trap holonomic quantum computers}

\author{V.I. Kuvshinov$^1$, A.V. Kuzmin$^2$ \\
Joint Institute for Power and Nuclear Research "Sosny" \\
220109 Krasina str., 99, Minsk, Belarus \\
$1$ - e-mail: v.kuvshinov@sosny.bas-net.by \\
$2$ - e-mails: avkuzmin@sosny.bas-net.by;
avkuzmin@dragon.bas-net.by.}

\date{}

\maketitle

\begin{abstract}
We consider one possible implementation of Hadamard gate for
optical and ion trap holonomic quantum computers. The expression
for its fidelity determining the gate stability with respect to
the errors in the single-mode squeezing parameter control is
analytically derived. We demonstrate by means of this expression
the cancellation of the squeezing control errors up to the fourth
order on their magnitude.
\end{abstract}

Holonomic quantum computation exploiting non-abelian geometrical
phases (holonomies) was primarily proposed in
Ref.~\cite{Zanardi1999} and was further worked out in
Ref.~\cite{Pachos2000}. Various implementations of holonomic
quantum computer (HQC) have been proposed recently. Namely, it was
suggested to realize the HQC within quantum optics (optical
HQC)~\cite{OHQC}. Laser beams in a non-linear Kerr medium were
used for this purpose. Two different sets of control devices could
be used in this case. The first one consists of one and two mode
displacing and squeezing devices. The second one includes SU(2)
interferometers. Squeezing and displacement of the vibrational
modes of the trapped ions were suggested to use for realization of
HQC in Ref.~\cite{ion}. This implementation of HQC is
mathematically similar to the first embodiment of the optical HQC
offered in Ref.~\cite{OHQC}. Particularly, expressions for the
adiabatic connection and holonomies are the same. Another proposed
implementation of HQC was HQC with neutral atoms in cavity
QED~\cite{NeutralHQC}. The coding space was spanned by the dark
states of the atom trapped in a cavity. Dynamics of the system was
governed by the generalized $\Lambda$-system Hamiltonian.
Mathematically similar semiconductor-based implementation of HQC
was proposed in Ref.~\cite{Semicond}. One-qubit gates were
realized in the framework of the same generalized $\Lambda$-system
as in Ref.~\cite{NeutralHQC}. However its physical implementation
exploits semiconductor excitons driven by sequences of laser
pulses~\cite{Semicond}. For the two-qubit gate the bi-excitonic
shift was used. The generalized $\Lambda$-system with different
Rabi frequencies parametrization was exploited recently for HQC
implemented by Rf-SQUIDs coupled through a microwave
cavity~\cite{SQUID}. One more solid state implementation of HQC
based on Stark effect was proposed in Ref.~\cite{Stark}.

Quantum computers including HQC are analog-type devices. Thus
unavoidable errors in the assignment of the classical control
parameters lead to an errors in quantum gates and in the case when
the tolerance of quantum computation is not large enough the
computation fails. This obstacle (inaccuracy) is also related to
the decoherence problem~\cite{Shor}. The effect of the errors
originated from the imperfect control of classical parameters was
studied for ${\bf CP}^n$ model of HQC in Ref.~\cite{CPNEr} where
the control-not and Hadamard gates were particularly considered.
Approach based on the non-abelian Stokes theorem~\cite{Simonov}
was proposed in our previous Letter~\cite{WePLA}. Namely, the
general expression for fidelity valid for arbitrary implementation
of HQC in the case of the single control error having the
arbitrary size was derived. Simple approximate formulae was found
in the small error limit. Adiabatic dynamics of a quantum system
coupled to a noisy classical control field was studied in
Ref.~\cite{Gaitan}. It was demonstrated that stochastic phase
shift arising in the off-diagonal elements of the system's density
matrix can cause decoherence. The investigation of the robustness
of non-abelian holonomic quantum gates with respect to parametric
noise due to stochastic fluctuations of the control parameters was
presented in Ref.~\cite{Robust}. In this work three stability
regimes were discriminated for HQC model with logical qubits given
by polarized excitonic states controlled by laser pulses. Noise
cancellation effect for simple quantum systems was considered in
Ref.~\cite{Cancel}. The decoherence of HQC was discussed, for
instance, in Refs.~\cite{Decoh, Refocus}. Berry phase in classical
fluctuating magnetic field was considered in Ref.~\cite{Palma}.
Fidelity decay rates for HQC interacting with the stochastic
environment were obtained recently in Ref.~\cite{Buiv}.

In this Letter we consider one-qubit gates for optical HQC and HQC
on trapped ions. The mathematical model based on squeezing and
displacing transformations of the qubit state is the same for both
these implementations (compare~\cite{OHQC} and~\cite{ion}). We
consider one possible implementation of Hadamard gate and
analytically derive the expression for its fidelity determining
the gate stability with respect to the errors in the single-mode
squeezing parameter control. We demonstrate the cancellation of
the control errors up to the fourth order on their magnitude.

Let us briefly remind some results concerning HQC. In holonomic
quantum computer non-abelian geometric phases (holonomies) are
used for implementation of unitary transformations (quantum gates)
in the subspace $C^N$ spanned on eigenvectors corresponding the
degenerate eigenvalue of parametric isospectral family of
Hamiltonians $F = \{H(\lambda) = U(\lambda) H_0
U(\lambda)^+\}_{\lambda \in M}$, where $U(\lambda)$ is
unitary~\cite{Zanardi1999}. The $\lambda$'s are the control
parameters and $M$ represents the space of the control parameters.
The subspace $C^N$ is called quantum code ($N$ is the dimension of
the degenerate computational subspace). Quantum gates are realized
when the control parameters are adiabatically driven along the
loops in the control manifold $M$. The unitary operator mapping
the initial state vector into the final one has the form
$\bigoplus_{l=1}^R e^{i \phi_l} \Gamma_{\gamma}(A_\mu^l)$, where
$l$ enumerates the energy levels of the system, $\phi_l$ is the
dynamical phase, $R$ is the number of different energy levels of
the system under consideration and the holonomy associated with
the loop $\gamma \in M$ is given by:
\begin{equation}\label{Hol}
  \Gamma_{\gamma}(A_\mu) = \hat{P} \exp{\int_{\gamma} A_{\mu} d \lambda_{\mu} }.
\end{equation}
Here $\hat{P}$ denotes the path ordering operator, $A_\mu $ is the
matrix valued adiabatic connection given by the
expression~\cite{Wilczek}:
\begin{equation}\label{A}
(A_\mu )_{mn} = <\varphi_m|U^+ \frac{\partial}{\partial
\lambda_\mu} U|\varphi_n>,
\end{equation}
where $|\varphi_k>, \quad k=\overline{1,N}$ are the constant basis
vectors of the corresponding eigenspace $C^N$, index $\mu$
enumerates the classical control parameters of the system.
Dynamical phase will be omitted bellow due to the suitable choice
of the zero energy level. We shall consider the single subspace
(no energy level crossings are assumed).

For optical holonomic quantum computer (as well as for ion trap
HQC) one-qubit unitary transformations are given as a sequence of
single-mode squeezing and displacing operations $U = D(\eta)
S(\nu)$. Here:
\begin{equation}\label{SD}
  \begin{array}{c}
  \medskip
  S(\nu) = \exp{\left(\nu a^+ a^+ - \overline{\nu}a a\right)},
  \bigskip \\
  D(\eta) = \exp{\left( \eta a^+ - \overline{\eta} a \right)}
  \end{array}
\end{equation}
denotes single-mode squeezing and displacing operators
respectively, $\nu = r_1 e^{i\theta_1}$ and $\eta = x+iy$ are
corresponding complex control parameters, $a$ and $a^+$ are
annihilation and creation operators. The line over the parameters
denotes complex conjugate quantities. The full set of the
connection as well as the field strength matrix components can be
found in Refs.~\cite{OHQC,ion}. In this Letter we consider loops
belonging to the planes $(x,r_1)|_{\theta_1 = 0}$ and
$(y,r_1)|_{\theta_1 = 0}$ only. Corresponding field strength
components are~\cite{OHQC}:
\begin{equation}\label{F}
  \begin{array}{c}
  \medskip
  F_{x r_1}|_{\theta_1 = 0} = -2i\sigma_y \exp{(-2r_1)},
  \bigskip \\
  F_{y r_1}|_{\theta_1 = 0} = -2i\sigma_x \exp{(2 r_1)}.
  \end{array}
\end{equation}
Here $\sigma_x$ and $\sigma_y$ are Pauli matrices. The
corresponding holonomies for the loops $C_I \in (x,r_1)_{\theta_1
= 0}$ and $C_{II} \in (y,r_1)_{\theta_1 = 0}$ are given
by~\cite{OHQC}:
\begin{equation}\label{Gam}
  \begin{array}{c}
  \medskip
  \Gamma (C_I) = \exp{(-i\sigma_y \Sigma_I)}, \quad \Sigma_I =
  \int_{S(C_I)} dx dr_1 2 e^{-2r_1},
  \bigskip \\
  \Gamma (C_{II}) = \exp{(-i\sigma_x \Sigma_{II})}, \quad \Sigma_{II} =
  \int_{S(C_{II})} dy dr_1 2 e^{2r_1},
  \end{array}
\end{equation}
where $S(C_{I,II})$ are the regions in the planes
$(x,r_1)|_{\theta_1 = 0}$ and $(y,r_1)|_{\theta_1 = 0}$ enclosed
by the loops $C_I$ and $C_{II}$.

Hadamard gate is widely used in various quantum algorithms, for
example in quantum Fourier transform, for more details
see~\cite{Shor2}. It is given as follows:
\begin{equation}\label{Hadamard}
  H_0 = \frac{1}{\sqrt{2}} \left(\begin{array}{cc}
    1 & 1 \\
    1 & -1 \
  \end{array}\right).
\end{equation}
We can obtain it by the two successive $y$ and $x$ rotations on
$\pi/4$ and $\pi/2$ respectively:
\begin{equation}\label{iH0}
  -i H_0 = \Gamma(C_{II})_{\Sigma_{II}=\pi/2} \Gamma(C_I)_{\Sigma_I =
  \pi/4}.
\end{equation}
The overall phase factor $(-i)$ is not essential for our purposes.
From the experimentalist point of view it is more convenient to
hold three control parameters fixed and adiabatically vary the
fourth parameter. Thus we consider rectangular loops $C_I$ and
$C_{II}$ with its sides to be parallel to the coordinate axes. For
$C_I$ these sides are given by the lines $r_1 = 0$, $x=b_x$, $r_1
= d_x$, $x=a_x$. Here $a_x$ and $b_x$ can be chosen arbitrary and
the lengths of the rectangle's sides parallel to the $x$ axis are
$l_x = b_x - a_x$. In the case of the gate considered $d_x$ is
given by the following expression:
\begin{equation}\label{dx}
  d_x = - \frac{1}{2} \ln{\left(1-\frac{\pi}{4l_x}\right)}.
\end{equation}
It immediately follows that $l_x > \pi/4$. In the same way we set
the loop $C_{II}$ as the rectangle composed by the lines $r_1 =
0$, $y=b_y$, $r_1 = d_y$, $y=a_y$ and find that:
\begin{equation}\label{dy}
  d_y = \frac{1}{2}\ln{\left( 1+\frac{\pi}{2l_y} \right)},
\end{equation}
where $l_y = b_y - a_y$. To hold zero value of the squeezing
control parameter is more easy from the experimental point of view
than non-zero one. One encounters problems when trying to keep
non-zero squeezing parameter and simultaneously adiabatically
change the displacing parameter as well as vise versa. In this
Letter we restrict ourselves by the errors in the single-mode
squeezing parameter control.

To take into account the errors in assignment of the single-mode
squeezing parameter $r_1$ we have to replace $d_x$ by $d_x +
\delta r_x (x)$ and $d_y$ by $d_y + \delta r_y (y)$. In this case
we obtain that parameters $\Sigma_I$ and $\Sigma_{II}$ entering
into the expressions~(\ref{Gam}) are replaced by:
\begin{equation}\label{Sigma}
  \begin{array}{c}
  {\displaystyle \medskip
  \Sigma_I^{\prime} = \Sigma_I +
  e^{-2d_x} \int_{a_x}^{b_x} dx \left( 1 - e^{-2\delta r_x}
  \right) = \Sigma_I + \delta \Sigma_I,}
  \bigskip \\
  {\displaystyle  \Sigma_{II}^{\prime} = \Sigma_{II} +
  e^{2d_y} \int_{a_y}^{b_y} dy \left( e^{2\delta r_y}-1
  \right) = \Sigma_{II} + \delta \Sigma_{II}.}
  \end{array}
\end{equation}
Therefore the perturbed Hadamard gate is given by the following
expression:
\begin{equation}\label{PertH}
  -i H = \Gamma(C_{II})|_{\Sigma_{II}^{\prime}}
  \Gamma(C_I)|_{\Sigma_I^{\prime}}.
\end{equation}
Using~(\ref{Gam}) and (\ref{Sigma}) we obtain:
\begin{equation}\label{LongH}
\begin{array}{c}
{\displaystyle \medskip
  -i H = - \frac{1}{\sqrt{2}} \left( \cos{\delta \Sigma_I} - \sin{\delta \Sigma_I} \right)
  \left( I \sin{\delta \Sigma_{II}} + i \sigma_x \cos{\delta \Sigma_{II}}
  \right)- }
  \bigskip
  \\
  {\displaystyle
  - \frac{i}{\sqrt{2}}\left( \cos{\delta \Sigma_I} + \sin{\delta \Sigma_I} \right)
  \left( \sigma_z \cos{\delta \Sigma_{II}} - \sigma_y \sin{\delta \Sigma_{II}}
  \right)},
  \end{array}
\end{equation}
where $\sigma_x$, $\sigma_y$ and $\sigma_z$ are Pauli matrixes and
$I$ is $2\times 2$ identity matrix. Fidelity of the Hadamard gate
determining the gate stability with respect to the errors in the
control of the single-mode squeezing parameter $r_1$ is:
\begin{equation}\label{fid}
  f_j = \sqrt{|<j|iH_0^+(-iH)|j>|^2}, \quad j=0,1.
\end{equation}
Here $|0>$ and $|1>$ are the basis vectors of the qubit.
Substituting expressions (\ref{Hadamard}) and (\ref{LongH}) into
formulae (\ref{fid}) we obtain:
\begin{equation}\label{fidel}
  f = |\cos{\delta \Sigma_I}|.
\end{equation}
Here we see that fidelity does not depend  on $j$ and $f_j \equiv
f$. As well it is evident that fidelity does not depend on errors
made in the $(y,r_1)_{\theta_1 = 0}$ plane. The reason is that the
corresponding $x$-rotation up to the overall phase factor is just
the classical not-gate.

Using expressions (\ref{Sigma}), (\ref{fidel}) and assuming that
the control error $\delta r_x (x)$ much less than unity for all
$x$ and have the zero average value at the line segment
$[a_x,b_x]$ we find:
\begin{equation}\label{fidelity}
  f \simeq \left|\cos{\left[<\delta r_x^2> \left(2l_x -
  \frac{\pi}{2}\right)\right]}\right| \simeq 1 - \left(<\delta r_x^2>\right)^2
  \left[ l_x \sqrt{2} - \frac{\pi}{2\sqrt{2}} \right]^2,
\end{equation}
where
\begin{equation}\label{r2}
  <\delta r_x^2> = \frac{1}{l_x} \int_{a_x}^{b_x} \delta r_x^2 (x)
  dx.
\end{equation}
Since $l_x > \pi/4$ fidelity $f=1$ at $l_x = \pi/4$ and equal or
less than unity for all over values of the parameters as it should
be. As well fidelity is equal to unity for $l_x^{(n)} = \pi/4 +
\pi n /(2<\delta r_x^2>)$ with integer $n > 0$. We believe that
this result stems from the fact that we have restricted ourselves
by the errors in the squeezing parameter control only. If we take
into account errors in assignment of the over control parameters,
especially fluctuations of displacing parameters $x$ and $y$ while
the squeezing parameter is being adiabatically changed, fidelity
of the Hadamard gate will be less than unity for all $l_x^{(n)},
\quad n>0$. However, there are reasons to believe that local
maxima at these points will still remain. To clarify it is the
task for the further investigations. As well we would like to note
the cancellation of the squeezing control errors up to the fourth
order on their magnitude that is obviously follows from the
expression (\ref{fidelity}) where we took into account that linear
terms were absent in the cosine Taylor expansion. Thus, in this
Letter we have analytically derived the expression for the
fidelity determining the Hadamard gate stability with respect to
the errors in the single-mode squeezing parameter control. We have
demonstrated the cancellation of the control errors up to the
fourth order on their magnitude under the restrictions and
conditions stated above.

\end{document}